\documentclass[a4paper,12pt,english]{article}

\usepackage{amsfonts,bm,amssymb,euscript,array,babel,cite}
\usepackage{amsmath,amsthm}
\usepackage[dvips]{epsfig}
\usepackage{slashed}
\usepackage{amsmath}



\newcommand{\be}{\begin{equation}} \newcommand{\ee}{\end{equation}}
\newcommand{\beq}{\begin{equation}} \newcommand{\eeq}{\end{equation}}
\newcommand{\beqa}{\begin{eqnarray}}
\newcommand{\eeqa}{\end{eqnarray}} \newcommand{\eq}[1]{(\ref{#1})}
\def\nn{\nonumber} \def\bea{\begin{eqnarray}} \def\eea{\end{eqnarray}}
\def\obar{\overline}

\newcommand{\barr}{\begin{array}}
\newcommand{\earr}{\end{array}}

\def\a{\alpha}  \def\b{\beta}
 \def\g{\gamma} 
 \def\d{\delta} 
 \def\e{\epsilon} 
\def\f{\phi} \def\F{\Phi}   
\def\l{\lambda}   
 \def\o{\omega}

      \def\cN{{\cal N}}

\def\R{{\mathbb R}}  
\def\Z{{\mathbb Z}} \def\one{\mbox{1 \kern-.59em {\rm l}}}

\def\bit{\begin{itemize}} \def\eit{\end{itemize}} 

\def\({\left(} \def\){\right)}

 \allowdisplaybreaks[3]

\begin{document}

\renewcommand{\title}[1]{\vspace{10mm}\noindent{\Large{\bf
#1}}\vspace{8mm}} \newcommand{\authors}[1]{\noindent{\large
#1}\vspace{5mm}} \newcommand{\address}[1]{{\itshape #1\vspace{2mm}}}

\begin{titlepage}
\begin{flushright}
CERN-PH-TH/2010-015\\
UWThPh-2010-2
\end{flushright}

\begin{center}

\title{ \Large Orbifolds, fuzzy spheres and chiral fermions}

\vskip 3mm

\authors{Athanasios {Chatzistavrakidis${}^{1,2}$},
Harold {Steinacker${}^3$},  \\[1ex]
 George {Zoupanos${}^{2,4}$}}

\vskip 3mm

\address{${}^1$ {\it Institute of Nuclear Physics,
NCSR  Demokritos,\\
GR-15310 Athens, Greece}\\[1ex] ${}^2$ Physics Department, National
Technical University,\\ Zografou Campus, GR-15780 Athens, Greece\\[1ex]
${}^3$Institut for Theoretical Physics, University of Vienna,\\
Boltzmanngasse 5, A-1090 Wien, Austria\\[1ex] ${}^4$ Theory Group, Physics Department\\
CERN, Geneva, Switzerland \\[3ex] E-mails:
{cthan@mail.ntua.gr, harold.steinacker@univie.ac.at, George.Zoupanos@cern.ch}}

\vskip 1.4cm

\textbf{Abstract}

\end{center}



Starting with a ${\cal N}=4$ supersymmetric Yang-Mills theory in four
dimensions with gauge group $SU(3N)$ we perform an orbifold projection 
leading to a ${\cal N}=1$ supersymmetric $SU(N)^3$ Yang-Mills theory with 
matter supermultiplets in bifundamental representations of the gauge 
group, which is chiral and anomaly free. Subsequently, we search for vacua 
of the projected theory which can be interpreted as spontaneously 
generated twisted fuzzy spheres. We show that by adding the appropriate 
soft supersymmetry breaking terms we can indeed reveal such vacua. Three 
cases are studied, where the gauge group is spontaneously broken further 
to the low-energy gauge groups $SU(4)\times SU(2)\times SU(2)$, $SU(4)^3$ 
and $SU(3)^3$. Such models behave in intermediate scales as 
higher-dimensional theories with a finite Kaluza-Klein tower, while their 
low-energy physics is governed by the corresponding zero-modes and 
exhibit chirality in the fermionic sector. The most interesting case from 
the phenomenological point of view turns out to be the $SU(3)^3$ unified 
theory, which has several interesting features such as (i) it can be 
promoted to a finite theory, (ii) it breaks further spontaneously first 
to the MSSM and then to $SU(3)\times U(1)_{em}$ due to its own scalar 
sector, i.e. without the need of additional superfields and (iii) the 
corresponding vacua lead to spontaneously generated fuzzy spheres.

\vskip 3mm

\begin{minipage}{14cm}%

\end{minipage}

\end{titlepage}

\tableofcontents


\section{Introduction}

The Standard Model (SM) is the most successful theory of elementary particle physics to date, bearing a remarkable agreement to the experimental data. However, the main limitation it confronts is the presence of a large number of free parameters in the theory. In order to reduce these parameters one is usually led to introduce more symmetry. Grand Unified Theories (GUTs) are very good examples of such a procedure \cite{Georgi:1974sy,Georgi:1974yf,Fritzsch:1974nn,Pati:1973rp}, where (approximate) gauge coupling unification can be achieved. Moreover, LEP data \cite{Amaldi:1991cn} seem to suggest that a further symmetry, namely ${\cal N} = 1$ global supersymmetry \cite{Dimopoulos:1981zb,Sakai:1981gr}, should also be required.

In addition to gauge coupling unification it is important to understand the origin of the Higgs and Yukawa sectors, which in the SM are introduced in an ad hoc manner and include most of its free parameters. A popular proposal is that gauge-Higgs-Yukawa unification can be achieved in higher dimensions. Pioneering studies in this direction were the Coset Space Dimensional Reduction (CSDR) \cite{Forgacs:1979zs,Kapetanakis:1992hf,Kubyshin:1989vd} and the Scherk-Schwarz reduction mechanism \cite{Scherk:1979zr}. In these frameworks the four-dimensional gauge and
Higgs fields are simply the surviving components of the gauge fields of a pure gauge theory defined in higher dimensions. Moreover in the CSDR the addition of fermions in the higher-dimensional gauge theory leads naturally to Yukawa couplings in four dimensions. A major achievement in this direction is the possibility to obtain chiral theories in four dimensions \cite{Manton:1981es}.

On the other hand, the theoretical efforts to establish a unified
description of all fundamental interactions including gravity led to
the development of superstring theory, which is consistent only in ten
dimensions. In addition, the heterotic string theory
\cite{Gross:1985fr} suggests that the gauge group in ten dimensions is
$E_8\times E_8$ or $SO(32)$, which are large enough to accommodate the
gauge group of the SM. The heterotic string theory is related via
certain dualities to the type II superstring theories, which include
$D$-branes. The discovery of $D$-branes paved a new road 
towards the unification of fundamental interactions \cite{Polchinski:1998rr}. 
After the discovery that stacks of branes naturally lead to non-abelian structures it was proposed that one could start with the SM or some extension of it in four dimensions and subsequently embed this local picture in a global context \cite{Aldazabal:2000sa}.

In order to bring in contact the superstring theory and the low energy
phenomenology it is crucial to dimensionally reduce the original
theory to four dimensions. A toroidal dimensional reduction from ten
to four dimensions leads to ${\cal N}=4$ supersymmetry in four
dimensions, which is not phenomenologically acceptable. The obvious
way to obtain ${\cal N}=1$ four-dimensional models, which might be
realistic, is to reduce the theory on suitable manifolds such as
Calabi-Yau manifolds \cite{Candelas:1985en} or manifolds with an
$SU(3)$-structure (see
e.g. \cite{Gauntlett:2003cy,LopesCardoso:2002hd}). It is worth noting
that using the CSDR scheme it is possible to obtain also the soft supersymmetry breaking sector of the four-dimensional theory \cite{Manousselis:2000aj}.

However, motivated by the celebrated duality between four-dimensional
${\cal N} = 4$ Supersymmetric $U(N)$ Yang-Mills (SYM) theory and Type IIB string theory on $AdS_5\times S^5$ \cite{Maldacena:1997re}, the authors of \cite{Kachru:1998ys} used orbifold techniques similar to \cite{Douglas:1997de,Douglas:1996sw} to break some of the supersymmetries. Considering different embeddings of a $\Z_3$ discrete group in the $R$-symmetry group of the ${\cal N}=4$ SYM theory and performing an orbifold projection of the original theory they determined ${\cal N}=0,1,2$ theories, i.e. with reduced supersymmetry. Moreover, the initial gauge group $SU(3N)$ (realised on $3N$ $D3$ branes) is broken down to $SU(N)^3$. 

Similar ideas have been developed also in the case where the extra dimensions are matrix approximations of smooth manifolds, i.e. fuzzy spaces. In \cite{Aschieri:2005wm} it was shown that starting with a higher-dimensional gauge theory and dimensionally reducing it over a fuzzy coset space several interesting features appear. Most importantly the theory is renormalizable in four as well as in higher dimensions. Moreover, non-abelian structures appear in four dimensions even if one starts with an abelian gauge theory in higher dimensions. However, the accommodation of chiral fermions in four dimensions turns out to be a difficult task in this context.

Motivated by the idea of deconstruction of dimensions \cite{Arkani-Hamed:2001ca}, 
the above approach was reversed in \cite{Aschieri:2006uw} in order to further justify the 
renormalizability of the theory and in an attempt to build chiral models in theories arising from 
fuzzy extra dimensions. Moreover, the reversed procedure gives the hope that not even the initial 
abelian gauge theory is necessary in higher dimensions, but the non-abelian gauge theory can emerge 
from fluctuations of the coordinates \cite{Steinacker:2003sd}. 
This approach consists in starting with a four-dimensional gauge theory with an appropriate content of scalar fields and a potential which can lead to vacua which are interpreted as dynamically generated fuzzy extra dimensions and they include a finite Kaluza-Klein tower of massive modes. The inclusion of fermions in such models showed that the best one could achieve so far is to obtain theories with mirror fermions in bifundamental representations of the low-energy gauge group \cite{Chatzistavrakidis:2009ix,Steinacker:2007ay}. Although mirror fermions do not exclude the possibility to make contact with phenomenology \cite{Maalampi:1988va}, it would be desirable to obtain exactly chiral fermions. 

The purpose of the present paper is to show that it is indeed possible to construct chiral unification models in this context upon introducing an additional geometrical structure, namely an orbifold structure as in \cite{Kachru:1998ys}. 
Our starting point is an orbifold projection of a
${\cal N}=4$ SYM theory, which is a
perfectly well-defined quantum field theory.
We will show that a twisted extra-dimensional
fuzzy sphere can arise dynamically in such orbifold models, breaking
the full $SU(N)^3$ gauge theory spontaneously down to
a chiral effective theory with unbroken gauge group such as e.g. $SU(3)^3$. 
Thus the scalar fields acquire a geometrical interpretation in terms of 
extra dimensions.
Moreover, the same type of fuzzy orbifolds can be used to break the
gauge symmetry spontaneously down to the Minimal Supersymmetric
Standard Model (MSSM) and furthermore to the $SU(3)_c\times U(1)_{\rm em}$. 
Thus fuzzy extra dimensions are realized in a chiral model which is not
only renormalizable but may even be finite and phenomenologically viable. 

Similar constructions have been studied in the framework of YM matrix models \cite{Grosse:2010zq}, 
which were however not phenomenologically viable. The mechanism we present here resolves the
 unrealistic features of these models and therefore it opens up the possibility to carry over 
our results to a theory describing dynamical noncommutative spacetime and gravity. 
Fuzzy spheres have been shown to appear as classical solutions in this framework in 
\cite{Iso:2001mg,Azuma:2004zq}, while orbifold techniques have been applied 
in the matrix model in \cite{Aoki:2002jt}.

The outline of the paper is the following. In section 2 the basics of
the projection of the ${\cal N}=4$ supersymmetric Yang-Mills theory
due to a $\Z_3$ orbifold is reviewed and the field content and the
superpotential of the projected theory are presented. In section 3 the
twisted fuzzy sphere is introduced and its relation to the ordinary
fuzzy sphere is presented.  Subsequently, the potential of the
orbifold projected theory is studied and a search for vacua which
develop dynamically spherical fuzzy extra dimensions is performed. It
is argued that such vacua can only be obtained when the appropriate
soft supersymmetry breaking terms are included in the potential. 
Having determined the general form of such vacua we proceed in section
4 to the study of specific anomaly-free models. Requiring that these
models are phenomenologically viable and focusing on the minimal
cases, we study the gauge groups $SU(4)\times SU(2)\times
SU(2),SU(4)^3$ and $SU(3)^3$. 
In section 5 the previous models are revisited and their spontaneous
breaking down to the MSSM and the $SU(3)_c\times U(1)_{\rm em}$ is
studied. The most interesting case from the phenomenological point of
view turns out to be the $SU(3)^3$, which can be promoted to a finite
theory 
and it has predictive power. Section 6 summarises our conclusions.

\section{$\Z_3$ orbifolds of ${\cal N}=4$ SYM}

In this section we review the basics of the $\Z_3$ orbifold projection of the ${\cal N}=4$ Supersymmetric Yang-Mills (SYM) theory \cite{Brink:1976bc}, in order to make the paper self-contained and to fix our notation. In particular we discuss the action of the discrete group on the various fields of the theory and the resulting superpotential of the projected theory. 

Before introducing the orbifold projection, the theory under consideration is the ${\cal N}=4$ supersymmetric $SU(3N)$ gauge theory\footnote{The gauge group is taken to be $SU(3N)$ for notational convenience as it will be clear in the following.}. This theory contains, in ${\cal N}=1$ language, a $SU(3N)$ vector supermultiplet and three adjoint chiral supermultiplets $\F^i,i=1,2,3$. The component fields are the $SU(3N)$ gauge bosons $A_{\mu},\mu=1,\dots ,4$, six adjoint real scalars\footnote{In the following we shall often work with the three complex scalars $\f^i,i=1,2,3$, which correspond to the complexification of the six real ones.} $\f^a,a=1,\dots, 6$, transforming as $\textbf{6}$ under the $SU(4)_R$ $R$-symmetry of the theory and four adjoint Weyl fermions $\psi^p,p=1,\dots ,4,$ transforming as $\textbf{4}$ under the $SU(4)_R$. The theory is defined on the Minkowski spacetime, whose coordinates are denoted as $x^{\mu},\mu=1,\dots,4$.

 In order to discuss orbifolds we have to consider the discrete group $\Z_3$ generically as a subgroup of $SU(4)_R$. There are three possibilities here, which have a direct impact on the amount of remnant supersymmetry \cite{Kachru:1998ys}. The first possibility is to embed the group $\Z_3$ maximally in $SU(4)_R$, in which case we are generically led to non-supersymmetric models. Secondly, the discrete group can be embedded in an $SU(3)$ subgroup of the full $R$-symmetry group, leading to $\cN=1$ supersymmetric models with $R$-symmetry $U(1)_R$. Specifically, this embedding can be viewed through the following decompositions of the vector and spinor of $SU(4)_R$ under $SU(3)\times U(1)_R$,
\bea 
SU(4)_R&\supset & SU(3)\times U(1)_R\nn\\
4 &=& 1_3 +3_{-1}, \nn\\
6&=& 3_2+\overline{3}_{-2}.
\eea
It is then clear that the spinor of $SU(4)_R$ decomposes into a singlet, which corresponds to the gaugino of the ${\cal N}=1$ theory and a triplet $\psi^i$, where the superpartners of the complex scalars are accommodated.
The last possibility is to embed $\Z_3$ in a specific $SU(2)$ subgroup of $SU(4)_R$, in which case there exist two surviving gaugini and therefore the remaining supersymmetry is ${\cal N}=2$.

Let us next discuss in more detail the case where $\Z_3$ is embedded in $SU(3)$, that leads to $\cN=1$ supersymmetric models. In order to proceed we consider a generator $g \in \Z_3$. This generator is conveniently labeled (see \cite{Douglas:1997de}) by three integers $\overrightarrow{a}\equiv (a_1,a_2,a_3)$ which satisfy the condition $a_1+a_2+a_3\equiv 0 \ \mbox{mod} \ 3$. This condition is equivalent to the statement that the discrete group is indeed embedded in $SU(3)$ and therefore it reflects the fact that $\cN=1$ supersymmetry is preserved \cite{Bailin:1999nk}. 

The $\Z_3$ acts non-trivially on the various fields of the theory depending on their transformation properties under the $R$-symmetry. 
The geometric action of the $\Z_3$ rotation on the gauge and the gaugino fields is trivial, since they are singlets under $SU(4)_R$. 
 On the other hand, the action of $\Z_3$ on the complex scalars is specified by the matrix \be \g(g)_{ij}=\d_{ij}\o^{a_i}, 
\ee where $\o=e^{\frac{2\pi i}{3}}$, while the corresponding action on the fermions $\psi^i$ is given by  \be \g(g)_{ij}=\d_{ij}\o^{b_i}, 
\ee where 
\bea b_1&=&-\frac{a_2+a_3-a_1}{2}, \nn\\
		 b_2&=&-\frac{a_1-a_2+a_3}{2}, \nn\\
			 b_3&=&-\frac{a_1+a_2-a_3}{2}.
				 \eea
In the case under study the three integers have the values $\overrightarrow{a}=(1,1,-2)$, which implies
$b_i = a_i$.

However, since the matter fields also transform non-trivially under the gauge group, the discrete group acts on their gauge indices too. 
 The action of this rotation can be described by the matrix
\be\label{g3}
\g_3 = \left(\begin{array}{ccc} \one_{N} & 0 & 0 \\
        0 & \omega\one_{N} & 0 \\
        0 & 0 & \omega^2\one_{N} \end{array}\right).
\ee
Let us note that in general the blocks of this matrix could have different dimensionality 
(see e.g. \cite{Lawrence:1998ja,Aldazabal:2000sa,Kiritsis:2003mc}),
However, anomaly freedom of the projected theory typically
requires that the dimension of the three blocks is the same
as will become obvious in the following. 
There is an interesting exception to this rule which will be discussed
in section \ref{sec:4222-model}. 

In order to derive the projected theory under the orbifold action, one has to keep the fields which are invariant under the combined action of the discrete group on the geometry and on the gauge indices \cite{Douglas:1997de}. For the gauge bosons the relevant projection is 
\be A_{\mu}=\g_3A_{\mu}\g_3^{-1}. \ee Therefore, in view of (\ref{g3}), the gauge group $SU(3N)$ of the original theory is broken down to $H=SU(N)\times SU(N)\times SU(N)$ in the projected theory.

For the complex scalars, which transform non-trivially both under the gauge group and the $R$-symmetry, the projection is \be\label{scalartransform} \phi^i= \o^{a_i}\g_3\phi^i\g_3^{-1}, \ee or, exhibiting the gauge indices which are denoted as $I,J$,
\be \f^i_{IJ}=\o^{I-J+a_i}\f^i_{IJ}. \ee This means that $J=I+a_i$ and therefore it is easy to see that the fields which survive the orbifold projection have the form $\f_{I,I+a_i}$ and they transform under the gauge group $H$ as
\be
\label{repsH} 3\cdot \biggl((N,\overline{N},1)+(\overline{N},1,N)+(1,N,\overline{N})\biggl). 
\ee 

For the fermions the situation is practically the same.
More specifically in this case the relevant projection is 
\be 
\psi^i= \o^{b_i}\g_3\psi^i\g_3^{-1} 
\ee 
or, equivalently,
\be \psi^i_{IJ}=\o^{I-J+b_i}\psi^i_{IJ}. 
\ee 
Then the surviving 
fermions have the form $\psi^i_{I,I+b_i}$ and they transform under 
$H$ in the representations (\ref{repsH}), exactly as the scalars. 
This is just another manifestation of the $\cN=1$ remnant supersymmetry. 
Moreover, the structure of the representations (\ref{repsH}) 
guarantees that the resulting theory does not suffer from any 
gauge anomalies\footnote{On the contrary, had we considered that 
the matrix (\ref{g3}) contained blocks of different dimensionality 
the projected theory would be anomalous and therefore additional 
sectors would be necessary in order to cancel the gauge anomalies.}. 

Let us next note two important features of the projected theory. First the fermions transform in chiral representations of the gauge group. Indeed, the representations (\ref{repsH}) are complex bifundamental ones, and their complex conjugates do not appear in the projected theory. Secondly, there are three fermionic generations in the theory. This is expected since as we noted before the theory contains three chiral supermultiplets under ${\cal N}=1$, leading to three generations.

Concerning the interactions among the fields of the projected theory, let us consider the superpotential of the $\cN=4$ supersymmetric Yang-Mills theory, which has the form \cite{Brink:1976bc}:
\be\label{spot1} W_{\cN=4} =\epsilon_{ijk}Tr(\F^i\F^j\F^k)
,
\ee where the three chiral superfields of the theory appear. 
Clearly, the superpotential after the orbifold projection has the same form but it encodes only the interactions among the surviving fields of the resulting ${\cal N}=1$ theory. Therefore it can be written as
\be\label{spot3} W_{{\cal N}=1}^{(proj)}=\sum_{I}\e_{ijk}\F^i_{I,I+a_i}\F^j_{I+a_i,I+a_i+a_j}\F^k_{I+a_i+a_j,I}
, \ee  
where the relation $a_1+a_2+a_3\equiv 0 \ \mbox{mod} \ 3$ was taken into account.

\section{Orbifolds for fuzzy spheres}

\subsection{Twisted fuzzy spheres}
\label{sec:twisted}

In the present section we introduce the "twisted fuzzy sphere" $\tilde S^2_N$, which is 
a variant of the ordinary fuzzy sphere \cite{Madore:1991bw} compatible with the orbifolding.
It is defined by the following relations
\bea
\label{twisted-vacuum} 
[\f^i,\f^j]&=&i\e_{ijk}(\f^k)^{\dagger}, \\ \f^i (\f^i)^{\dagger} &=&  R^2,
\eea
where $(\f^i)^{\dagger}$ denotes hermitean conjugation of the complex scalar field $\f^i$
and $[ R^2,\f^i] = 0$.
The relation (\ref{twisted-vacuum}) is compatible with the $\Z_3$ group action (\ref{scalartransform}), in contrast to the usual fuzzy sphere.
Nevertheless the above relations are closely related to a fuzzy sphere.
This can be seen by considering the untwisted fields $\tilde \f_i$, defined by
\be
\f^i = \Omega\,\tilde\f^i , 
\label{twisted-fields}
\ee
for some $\Omega\ne 1$ which satisfies
\be
\Omega^3 = 1, \quad [\Omega,\f^i] = 0, \quad \Omega^\dagger = \Omega^{-1}
\label{cond-1}
\ee
and\footnote{Here $[\Omega,\f^i]$ is understood before the orbifolding.}
\be
(\tilde\f^i)^\dagger = \tilde\f^i . \quad\mbox{i.e.}\,\,(\f^i)^\dagger = \Omega \f^i.
\label{cond-2}
\ee
Then \eq{twisted-vacuum} reduces to the ordinary fuzzy sphere relation
\be
\,[\tilde\f^i,\tilde\f^j] = i\e_{ijk}\tilde\f^k,
\label{fuzzy-transf} 
\ee
generated by $\tilde\f^i$, as well as to the relation
\be
\tilde \f^i\tilde \f^i = R^2.
\ee
This justifies to call the noncommutative space generated by $\f^i$ a twisted fuzzy sphere.
It is remarkable 
that this construction is possible only for $\Z_3$ and for no other
$\Z_n$, thus providing a justification for our choice of orbifold group.

Now let us discuss two different realizations of this twisted fuzzy sphere. 
The most obvious solution of \eq{twisted-vacuum} is given by
$\Omega = \omega$ and 
\be\label{twisted-vacuum-sol}
\phi^i = \omega \,\lambda^i_{(3N)}
,\ee where 
 $\lambda^i_{(3N)}$ denote the generators of $SU(2)$ in the 
$3N$-dimensional irreducible representation.

A second realization of a twisted fuzzy sphere (\ref{twisted-vacuum}) is given by
\be
\label{solution1} 
\phi^i = \Omega\, (\one_3\otimes\lambda^i_{(N)}),
\ee where the matrix $\Omega$ is given by
\be
\Omega = \Omega_3 \otimes \one_N, \quad
\Omega_3 = \begin{pmatrix}
		0 & 1 & 0 \\
		0 & 0 & 1 \\
		1 & 0 & 0 \\
\end{pmatrix}, \qquad \Omega^3 = \one. 
\label{omega-large}
\ee 
The transformation 
$\f^i = \Omega\,\tilde\f^i$ 
\eq{twisted-fields} relates the 
"off-diagonal" orbifold sectors \eq{repsH} to block-diagonal configurations as follows,

\be\label{transuntwist}
\f^i = \begin{pmatrix}
		0 & (\l^i_{(N)})_{(N,\overline N,1)} & 0 \\
		0 & 0 & (\l^i_{(N)})_{(1,N,\overline N)} \\
		(\l^i_{(N)})_{(\overline N,1,N)} & 0 & 0 \\
\end{pmatrix}
\quad = \quad \Omega \begin{pmatrix}
		\l^i_{(N)} & 0 & 0\\
		 0 & \l^i_{(N)} & 0 \\
		0 & 0 & \l^i_{(N)}  \\
\end{pmatrix}.
\ee We observe that the untwisted fields $\tilde \f^i$, which generate the fuzzy sphere, acquire a block-diagonal form. Each one of these blocks satisfies separately the fuzzy sphere relation (\ref{fuzzy-transf}) and therefore it is natural to reinterpret this configuration as three fuzzy spheres of fuzziness $N$. 
The solution $\f^i$ can thus
be interpreted as twisted configuration of three fuzzy spheres
compatible with the orbifolding. 
Further insight can be gained by noting that $\Omega_3$ can be diagonalized
as $\Omega_3 = U^{-1} {\rm diag}(1,\omega,\omega^2) U$,
so that $\f^i$ can be interpreted as a combination of 
two basic solutions \eq{twisted-vacuum-sol} and an untwisted fuzzy sphere.

The solution (\ref{solution1}) breaks completely the gauge symmetry $SU(N)^3$. 
This geometrical interpretation 
is helpful to understand the fluctuations 
around these fuzzy orbifolds. However, for our purposes it will be useful to consider solutions which do not break the $SU(N)^3$ gauge symmetry completely but they break it down to a smaller gauge group. We shall study such solutions in the following paragraph and present specific applications in the sections 4 and 5.

\subsection{Dynamical generation of twisted fuzzy spheres}

Let us now show how the above geometries can arise as a vacuum solution of the 
field theory which was considered in section 2. As it was previously described, the superpotential of the theory after the orbifold projection has the form (\ref{spot3}). Therefore one can easily read off the corresponding potential, which is\footnote{Here we restrict to the scalar sector, since this is the relevant one for the search of fuzzy sphere vacua. Moreover, the gauge indices are suppressed.}
\be 
V_{{\cal N}=1}^{(proj)}(\f)= \frac 14 Tr([\f^i,\f^j]^\dagger [\f^i,\f^j]), 
\ee 
where $\f^i$ denotes the scalar component of the superfield $\F^i$. The minimum of this potential is obtained for vanishing vevs of the fields. It is therefore obvious that vacua with the geometry of fuzzy spheres cannot be obtained.

Searching for fuzzy sphere vacua would mean looking for a minimum which satisfies
\be 
[\f^i,\f^j]=i\e_{ijk}\f^k, 
\ee
which however is not compatible with the orbifold projection described in sections 2 and 3.1. However, a different minimum that is compatible with the orbifold has the form (\ref{twisted-vacuum}) as it was argued in the previous section.

Clearly, such a minimum calls for the following modifications in the theory. First of all, we have to add ${\cal N}=1$ soft supersymmetry breaking 
(SSB) terms of the form\footnote{Here we present a set of scalar SSB terms. However, there exist of 
course other soft terms such as $\frac 12 M\l\l$, where $\l$ is the gaugino and $M$ its mass, 
which has to be included in the full SSB sector \cite{Djouadi:2005gj}.}
\be\label{soft} 
V_{SSB}=\frac 12 \sum_i m^2_i\, {\f^i}^{\dagger}\f^i+\frac 12 \sum_{i,j,k}\, h_{ijk}\f^i\f^j\f^k+h.c., 
\ee 
where $h_{ijk}$ vanishes unless $i+j+k\equiv 0\mod 3$.
Of course
a set of SSB terms in the potential is necessary anyway in order for the theory to have a 
chance to become realistic, see e.g. \cite{Djouadi:2005gj}. After the addition of these soft terms as well as of the $D$-terms the 
full potential of the theory becomes 
\be\label{potential1} 
V=V_{{\cal N}=1}^{(proj)}+V_{SSB}+V_{D}, 
\ee
where $V_{D}=\frac 12 D^2=\frac 12 D^{I}D_{I}$ includes the $D$-terms of the theory. 
These $D$-terms have the form $D^{I}=\f_i^{\dagger}T^{I}\f^i$, where $T^I$ are the generators 
of the representation of the corresponding chiral multiplets. Let us note that in $V_D$ apart 
from the summation over the indices $i$ (labelling the chiral supermultiplets) and $I$ 
(the gauge group index), a summation over the three gauge group factors is also implied.

In order to allow for twisted
fuzzy sphere vacua, we now make the choice $h_{ijk}=\epsilon_{ijk}$ and $m_i^2=1$. 
A more general possibility will be investigated in section 5. Then the potential 
(\ref{potential1}) can be brought in the form
\be\label{potnewform} 
V=\frac 14 (F^{ij})^{\dag}F^{ij} \,\, + V_D, 
\ee
where we have defined
\be
\label{fieldstrength} 
F^{ij}=[\f^i,\f^j]-i\e^{ijk}(\f^k)^{\dagger}.
\ee
The first term of the potential is positive definite, and vanishes if the 
relation (\ref{twisted-vacuum}) holds. Therefore the global minimum of the potential
is realized by a twisted fuzzy sphere $\tilde S^2_N$ (14), 
at least for a suitable range of parameters 
in the potential. The quartic term $V_D$ will typically only modify
its radius, as in the case of the ordinary fuzzy sphere
\cite{Steinacker:2003sd,Aschieri:2006uw}. This vacuum will be studied in more detail below. 
 The expression (\ref{fieldstrength}) will be interpreted in the 
following as the field strength on the spontaneously generated fuzzy extra dimensions. 

In general, the potential 
may have several different local minima, 
which may be given e.g. by twisted fuzzy spheres with various radii; we will not discuss 
possible meta-stable vacua or phase-transitions here.

We have to specify the embedding of the $SU(2)$ corresponding to the fuzzy sphere structure of the vacuum into the $SU(4)_R$ $R$-symmetry 
of the original ${\cal N}=4$ SYM theory. Few group theory remarks are crucial in order to make the situation clear. One can easily check the following structure in $SU(4)_R$ and its maximal subgroups,
\bea 
SU(4)_R && \supset SU(2)^I\times SU(2)^{II}\times U(1) \nn\\
&& ~~~~~~~~~~~~~~\cup ~~~~~~~~~~~~~~~~\cap \nn\\
SU(4)_R&&\supset ~~~~~~~SU(2)^{\a}~~~\times ~~~SU(2)^{\b},
\eea
where $SU(2)^{\a}$ is the diagonal subgroup of the $SU(2)^I\times SU(2)^{II}$. Moreover, the diagonal subgroup of $SU(2)^{\a}\times SU(2)^{\b}$ is the $SU(2)^A$ that corresponds to the maximal embedding in $SU(3)$ according to the decomposition $SU(4)_R\supset SU(3)\times U(1)_R\supset SU(2)^A\times U(1)_R$. Clearly, in the maximal embedding it holds that 
\bea SU(3)&\supset &SU(2)^A\nn\\
3&=&3. \eea

Let us now study further the vacuum and its geometric interpretation. 
The scalar fields $\phi^i$ are governed by the potential (\ref{potential1}),
which includes the F- and D-terms as well as the SSB terms.
Under suitable conditions, 
this potential clearly has a twisted fuzzy sphere
solution 
\be
\phi^i = \Omega\, \biggl(\one_3\otimes(\lambda^i_{(N-n)} \oplus 0_{n})\biggl),
\label{twistedfuzzys2-2}
\ee
where $0_{n}$ denotes the $n\times n$ matrix with vanishing entries. 
The gauge symmetry is broken from $SU(N)^3$ down to $SU(n)^3$. 
This vacuum should be interpreted as $\R^4 \times \tilde S^2_N$
with a twisted 
fuzzy sphere in the $\phi_i$ coordinates.

In order to understand the fluctuations of the scalar fields around 
this vacuum, the transformation $\phi^i = \Omega\tilde\phi^i$ is useful.
Fluctuations around the ordinary fuzzy sphere
$S^2_N$ are known to describe gauge and scalar
fields on $S^2_N$ \cite{Madore:2000en,Steinacker:2003sd}, 
and in particular they all become massive 
from the point of view of $\R^4$.
We have seen in 
(\ref{transuntwist}) that the twisted sphere $\tilde S^2_N$ is mapped by $\Omega$ into three fuzzy spheres $\tilde \f^i$ 
embedded in the diagonal 
$N\times N$ blocks of the original $3N\times 3N$ matrix. 
Therefore all fluctuations can be understood as fields on the 
three diagonally embedded untwisted fuzzy spheres:
\be
\tilde \phi^i = \lambda^i_{(N)} + A^i ,
\ee
and the field strength \eq{fieldstrength} reduces to the field strength on a 
fuzzy sphere 
\be
F^{ij}=[\f^i,\f^j]-i\e^{ijk}(\f^k)^{\dagger} 
= \Omega^2 ([\tilde\f^i,\tilde\f^j]-i\e^{ijk}\tilde\f^k)
\ee
as long as \eq{cond-1} and \eq{cond-2} hold.
The vacuum can thus be interpreted at intermediate energy scales
as $\R^4 \times S^2_N$
with three (untwisted) fuzzy spheres 
in the $\tilde\phi_i^L$  coordinates.
Moreover, due to the orbifolding condition 
there are no off-diagonal components relating these different 
spheres. It now follows as in \cite{Aschieri:2006uw,Steinacker:2007ay}
that the gauge fields and fermions can be 
decomposed into Kaluza-Klein towers of massive modes
on $S^2_N$ resp. $\tilde S^2_N$
due to the Higgs effect, 
as well as a massless sector which will be elaborated below.

\section{Chiral models from the fuzzy orbifold}

In this section we discuss three particular models which can be constructed in the above context. 
In all cases we start by considering the ${\cal N}=4$ SYM theory in four dimensions with gauge group $SU(3N)$. As we have already mentioned this theory contains, in ${\cal N}=1$ language, an $SU(3N)$ vector supermultiplet and three adjoint chiral supermultiplets $\Phi^i$ with superpotential 
\be W_{{\cal N}=4}=\epsilon_{ijk}Tr(\F^i\F^j\F^k). 
\ee 

Subsequently we choose the discrete group $\Z_3$ and embed it in the $SU(3)$ part of the $R$-symmetry. Performing the orbifold projection, as it was described in section 2, we obtain an ${\cal N}=1$ theory with vectors in $SU(N)^3$ and complex scalars and fermions in chiral representations of the gauge group. In particular, according to (\ref{repsH}), there are three families, each transforming under the gauge group $H$ as 
\be\label{repsHmodel} (N,\overline{N},1)+(\overline{N},1,N)+(1,N,\overline{N}). \ee 
Moreover, the superpotential takes the form (\ref{spot3}). The difference between the models lies in the next step of the construction, where the gauge group $SU(N)^3$ will be broken spontaneously to a unification group. The minimal cases which satisfy the requirement of anomaly freedom are the gauge groups $SU(4)\times SU(2)\times SU(2), SU(4)^3$ and $SU(3)^3$.

\subsection{A $SU(4)_c\times SU(2)_L\times SU(2)_R$ model}
\label{sec:4222-model}

In order to obtain the Pati-Salam gauge group $SU(4)_c\times SU(2)_L\times SU(2)_R$ \cite{Pati:1974yy}, we decompose the integer $N$ in two different ways, namely as \bea N&=&n_1+4, \nn\\ N&=&n_2+2. \eea Then we consider the following regular embeddings,  
\bea SU(N)&\supset&SU(n_1)\times SU(4)\times U(1), \nn\\
SU(N)&\supset&SU(n_2)\times SU(2)\times U(1). 
\eea 

The full gauge group is accordingly decomposed as 
\be SU(N)^3\supset SU(n_1)\times SU(4)\times SU(n_2)\times SU(2)\times SU(n_2)\times SU(2)\times U(1)^3. \ee
Performing a shuffling of the group factors and ignoring the $U(1)$s{\footnote{These may be anomalous and become massive by the Green-Schwarz mechanism and therefore they decouple at low energies \cite{Lawrence:1998ja}.}} it is easy to see that the original representations (\ref{repsH}) are decomposed as follows,
\bea && SU(n_1)\times SU(n_2)\times SU(n_2)\times SU(4)\times SU(2)\times SU(2) \nn\\
&& (n_1,\obar{n}_2,1;1,1,1)+(1,n_2,\obar{n}_2;1,1,1)+(\obar{n}_1,1,n_2;1,1,1)+\nn\\
&& +(1,1,1;4,2,1)+(1,1,1;1,2,2)+(1,1,1;\obar{4},1,2)+\nn\\
&& +(n_1,1,1;1,2,1)+(1,n_2,1;1,1,2)+(1,1,n_2;\obar{4},1,1)+\nn\\
&& +(\obar{n}_1,1,1;1,1,2)+(1,\obar{n}_2,1;4,1,1)+(1,1,\obar{n}_2;1,2,1). \eea
First of all it is important to note that the theory is anomaly free. This is merely due to the special feature of $SU(2)$, where the fundamental representation is self-conjugate. Therefore, although the structure involves a product of different gauge groups, it is still not anomalous.

Now utilizing the mechanism of section 3, fuzzy extra dimensions can be dynamically generated and the unbroken gauge group at low-energies is $SU(4)_c\times SU(2)_L\times SU(2)_R$, with fields transforming under the representations 
\bea
\label{repsfinal2} & SU(4)\times SU(2)\times SU(2) \nn\\ 
		& 3\cdot \biggl((4,2,1)+(\overline{4},1,2)+(1,2,2)\biggl). 
\eea
This is realized by the following vacuum
\be
\phi^i = \Omega\,\biggl(0_2 \oplus \one_3\otimes (\lambda^i_{(N-2)} \oplus 0_{2})\biggl),
\qquad
\Omega = \begin{pmatrix}
		\one_2 & 0  \\
		0 & \Omega_3\otimes\one_{N}
\end{pmatrix}
\label{twistedfuzzys2-4}
\ee
interpreted in terms of twisted fuzzy spheres,
where $\Omega_3$ is defined in \eq{omega-large}.

Then the quarks and leptons of the SM fit in these representations. For example, the first generation is represented as
\bea\label{quarksleptons1}
f\sim(4,2,1) &=&  \left(\begin{array}{cc} d_L^1 & u_L^1 \\ d_L^2 & u_L^2 \\ d^3_L & u^3_L \\ e_L & \nu_L
\end{array}\right) ,  \nn\\
f^c\sim(\bar 4,1,2) &=&  \left(\begin{array}{cccc} d^{1 c}_L & d^{2 c}_L & 
d^{3 c}_L & e_L^c
\\ 
u^{1 c}_L & u^{2 c}_L & u^{3 c}_L & \nu^c_L \end{array}\right), 
\eea and similarly for the other two generations. Moreover, the $h\sim(1,2,2)$ representation involves the Higgses and the Higgsini.

\subsection{A $SU(4)_c\times SU(4)_L\times SU(4)_R$ model}

A further possibility is the gauge group $SU(4)_c\times SU(4)_L\times
SU(4)_R$, where $SU(4)_c$ is again the Pati-Salam colour gauge group. 
This gauge group can be obtained by decomposing $N$ as 
\be N=n+4, 
\ee 
leading to the decomposition of $SU(N)^3$ to $SU(n)^3\times SU(4)^3$ with particle content
\bea && SU(n)\times SU(n)\times SU(n)\times SU(4)\times SU(4)\times SU(4) \nn\\
&& (n,\obar{n},1;1,1,1)+(1,n,\obar{n};1,1,1)+(\obar{n},1,n;1,1,1)+\nn\\
&& +(1,1,1;4,\obar{4},1)+(1,1,1;1,4,\obar{4})+(1,1,1;\obar{4},1,4)+\nn\\
&& +(n,1,1;1,\obar{4},1)+(1,n,1;1,1,\obar{4})+(1,1,n;\obar{4},1,1)+\nn\\
&& +(\obar{n},1,1;1,1,4)+(1,\obar{n},1;4,1,1)+(1,1,\obar{n};1,4,1).
 \eea

This is realized by the following vacuum, interpreted  
in terms of twisted fuzzy spheres $\tilde S^2_{N-4}$ 
as in \eq{twistedfuzzys2-2}:
\be
\phi^i = \Omega\,\biggl(\one_3\otimes (\lambda^i_{(N-4)} \oplus 0_{4})\biggl),
\label{twistedfuzzys2-4}
\ee
where $\Omega$ is defined in \eq{omega-large}.
Decomposing $SU(N) \supset SU(n)\times SU(4)\times U(1)$,
the gauge group is broken to $SU(4)^3$, and the low-energy field 
content
is 
\bea\label{repsfinal3} & SU(4)\times SU(4)\times SU(4) \nn\\ 
		& 3\cdot \biggl((4,\overline{4},1)+(\overline{4},1,4)+(1,4,\overline{4})\biggl). \eea
This case has been examined originally in \cite{Ibanez:1998xn} and from a phenomenological viewpoint in \cite{Ma:2004mi}. The quarks and leptons of the first family should transform as
\begin{equation}
f = \left(\begin{array}{cccc} d & u & y & x \cr d & u & y & x \cr d & u & y & x \cr 
e & \nu & a & v \end{array}\right) \sim (4,\overline 4,1), ~~~ f^c = \left(\begin{array}{cccc} d^c & d^c & d^c & e^c 
\cr u^c & u^c & u^c & \nu^c \cr y^c & y^c & y^c & a^c \cr x^c & x^c & x^c & 
v^c \end{array}\right)\sim (\overline 4,1,4).
\label{quarksleptons2}
\end{equation}
Clearly, there have to be new heavy quarks and leptons and in addition the supermultiplet $h\sim(1,4,\overline{4})$ still has to be considered.

A very interesting feature which we would like to point out here is that the one-loop $\beta$-function coefficient in the renormalization group equation of each $SU(4)$ gauge coupling is given by 
\be b=\biggl(-\frac{11}{3}+\frac 23\biggl)\cdot3+n_{f}\biggl(\frac 23+\frac 13\biggl)\cdot\frac 12 \cdot2\cdot 3, \ee which for the present case of $n_f=3$ copies of the supermultiplet (\ref{repsfinal3}) results in \be b=0.\ee
Therefore, we observe that the existence of three families of quarks and leptons leads to one of the necessary conditions for a finite field theory. Let us mention that this is a general feature of models with a $SU(N)^k$ gauge group, independently of the values of $N$ and $k$ \cite{Ma:2004mi}. Therefore it also holds in the following case of $SU(3)^3$.

\subsection{A $SU(3)_c\times SU(3)_L\times SU(3)_R$ model}

Let us now turn to another possibility, the trinification group $SU(3)_c\times SU(3)_L\times SU(3)_R$ \cite{Glashow:1984gc,Rizov:1981dp}, which was also studied in \cite{Leontaris:2005ax,Lazarides:1993uw,Babu:1985gi,Ma:2004mi,Heinemeyer:2009zs}. In the present case we consider the following picture. Let us decompose the integer $N$ as \be N=n+3. \ee Subsequently, let us consider the regular embedding
\be\label{dec3} SU(N)\supset SU(n)\times SU(3)\times U(1). \ee  Then the relevant embedding for the full gauge group is
\be SU(N)^3\supset SU(n)\times SU(3)\times SU(n)\times SU(3)\times SU(n)\times SU(3)\times U(1)^3. \ee The three $U(1)$ factors decouple from the low-energy sector of the theory, as it was mentioned above. The representations (\ref{repsHmodel}) are then decomposed accordingly (notice the shuffling in the group factors),
\bea \label{repsall}
&& SU(n)\times SU(n)\times SU(n)\times SU(3)\times SU(3)\times SU(3)\nn\\
&& (n,\obar{n},1;1,1,1)+(1,n,\obar{n};1,1,1)+(\obar{n},1,n;1,1,1)+\nn\\
&& +(1,1,1;3,\obar{3},1)+(1,1,1;1,3,\obar{3})+(1,1,1;\obar{3},1,3)+\nn\\
&& +(n,1,1;1,\obar{3},1)+(1,n,1;1,1,\obar{3})+(1,1,n;\obar{3},1,1)+\nn\\
&& +(\obar{n},1,1;1,1,3)+(1,\obar{n},1;3,1,1)+(1,1,\obar{n};1,3,1).
\eea
This is realized by the following vacuum, interpreted  
in terms of twisted fuzzy spheres $\tilde S^2_{N-3}$ 
as in \eq{twistedfuzzys2-2}:
\be
\phi^i = \Omega\,[\one_3\otimes (\lambda^i_{(N-3)} \oplus 0_{3})].
\label{twistedfuzzys2-4}
\ee 
Considering the decomposition (\ref{dec3}),
the gauge group is broken
to $K = SU(3)^3$. 
Finally, the surviving fields under the unbroken gauge group $K$ transform in the following representations,
\bea\label{repsfinal} & SU(3)\times SU(3)\times SU(3) \nn\\ 
		& 3\cdot \biggl((3,\overline{3},1)+(\overline{3},1,3)+(1,3,\overline{3})\biggl). \eea
These are the desired chiral representations of the unification group $SU(3)_c\times SU(3)_L\times SU(3)_R$. 
The quarks of the first family transform under the gauge group as
\begin{equation}\label{quarks3}
q = \left(\begin{array}{ccc} d & u & h \cr d & u & h \cr d & u & h \end{array}\right)\sim (3,\overline 3,1), ~~~ 
q^c =\left(\begin{array}{ccc} d^c & d^c & d^c \cr u^c & u^c & u^c \cr h^c & h^c & h^c \end{array}\right) 
\sim (\overline 3,1,3),
\end{equation}
and the leptons transform as
\begin{equation}\label{leptons3}
\lambda = \left(\begin{array}{ccc} N & E^c & \nu \cr E & N^c & e \cr \nu^c & e^c & S  \end{array}\right) 
\sim (1,3,\overline 3).
\end{equation}
Similarly, the corresponding matrices for the quarks and leptons of the other two families can be written down.

\subsection{A closer look at the masses}

A vital issue of our construction is whether there exist massless and massive modes at the same time. Clearly we need both of these sets; the massless modes in order to obtain chiral fermions and the massive modes in order to reproduce the Kaluza-Klein tower and provide undoubtful justification that the theory develops fuzzy extra dimensions.

A way to see this through the embeddings we presented before is the following. Let us work out the case of $SU(3)^3$, since the same arguments apply to the other two cases as well. Under the final gauge group $SU(3)^3$ the fermions transform in the representations (\ref{repsfinal}), hence they are chiral. Therefore they remain massless since they are protected by chiral symmetry. 

On the other hand, looking at (\ref{repsall}) we can make two crucial observations. First of all, it becomes clear from the vacuum solution (\ref{twistedfuzzys2-4}) that the scalar fields which acquire vevs in this vacuum are the following,
\be \langle(n,\obar{n},1;1,1,1)\rangle, \langle(1,n,\obar{n};1,1,1)\rangle, \langle(\obar{n},1,n;1,1,1)\rangle.
\ee
Then all the fermions, apart from the chirally protected ones, obtain masses, since we can form the invariants
\bea \label{inv1} && (1,n,\obar{n};1,1,1)\langle(n,\obar{n},1;1,1,1)\rangle (\obar{n},1,n;1,1,1)\qquad \mbox{+ cyclic permutations},\\  \label{inv2} &&  (\obar{n},1,1;1,1,3)\langle(n,\obar{n},1;1,1,1)\rangle (1,n,1;1,1,\obar{3}) \qquad \mbox{etc.,}
\eea 
and the corresponding ones for all the other fermions. In these invariants the field in the middle is the scalar field which acquires the vev \eq{twistedfuzzys2-4}, 
while the other two are fermions, i.e. the invariants are trilinear Yukawa terms and they are responsible for the fermion masses after the spontaneous symmetry breaking. Therefore a finite Kaluza-Klein tower of massive fermionic modes appears, 
consistent with the interpretation of the vacuum (\ref{twistedfuzzys2-4}) as a higher-dimensional theory with spontaneously generated fuzzy extra dimensions. In particular, the fluctuations from this vacuum correspond to the internal components of the higher dimensional gauge field. Also, as far as the fermions transforming as $(1,1,1;3,\overline 3,1),(1,1,1;\overline 3,1,3)$ and $(1,1,1;1,3,\overline 3)$ are concerned, obviously there does not exist any trilinear invariant that they could form with one of the scalar fields which acquire a vev. Therefore, as it was already mentioned, they remain massless and they are the chiral fermions of the model.

Finally, it is worth noting that in (\ref{inv2}) the "internal" structure and the "observable", low-energy structure appear mixed and therefore these Kaluza-Klein fermion masses may have an effect on the $SU(3)^3$ 
phenomenology \cite{Dienes:1998vh,Ghilencea:1998st,Kubo:1999ua,Kobayashi:1998ye}.

\section{Fuzzy breaking and realistic models}

In this section we discuss another possible application of the fuzzy orbifold construction which was presented in section 3. The three models presented in section 4 are revisited and the orbifold projection is utilized to study their spontaneous breaking down to the MSSM and the $SU(3)_c\times U(1)_{em}$. It is important to note that we shall focus only on symmetry breaking patterns where additional superfields are not introduced, namely the models are broken spontaneously due to their own scalar sector. 

In particular, instead of starting with a $SU(3N)$ gauge theory with a large $N$, we can start with smaller gauge groups, in particular $SU(8),SU(12)$ and $SU(9)$, in order to obtain the models $SU(4)\times SU(2)\times SU(2), SU(4)^3$ and $SU(3)^3$ respectively after orbifolding. Therefore the initial set-up 
consists of the ${\cal N}=4$ SYM theory with gauge fields in one of the above gauge groups. Subsequently a $\Z_3$ orbifold projection is performed in the spirit of section 2.

Alternatively, this procedure may be viewed as a second step of the constructions which were presented in section 4. Indeed, if such a view is adopted, after the large-N symmetry breaking, the models presented in the previous section are obtained. They involve a superpotential and the corresponding soft supersymmetry breaking terms. Part of the SSB sector is naturally inherited from the corresponding one in the large-N models, namely it is already contained in the expression (\ref{soft}) for suitable $h_{ijk}$. This fact justifies further the use of the same technique, the spontaneous generation of twisted fuzzy spheres, in order to achieve the spontaneous symmetry breaking down to the MSSM and subsequently to the $SU(3)\times U(1)_{em}$.

\subsection{$SU(4)\times SU(2)\times SU(2)$ model}

The Pati-Salam gauge group can be obtained when the initial gauge group is $SU(8)$. 
We perform an orbifold projection of the ${\cal N}=4$, $SU(8)$ SYM theory such that the $\g_3$ of eq. (\ref{g3}) becomes
\be \g_3 = \left(\begin{array}{ccc} \one_{4} & 0 & 0 \\
        0 & \omega\one_{2} & 0 \\
        0 & 0 & \omega^2\one_{2} \end{array}\right). \ee
Then, according to the rules of section 2, the gauge group breaks down to $SU(4)\times SU(2)\times SU(2)$, with three chiral supermultiplets transforming as
\bea & SU(4)\times SU(2)\times SU(2) \nn\\
 & 3\cdot \biggl((4,2,1)+(\overline{4},1,2)+(1,2,2)\biggl). 
\eea
The quarks and leptons are accommodated in these representations as in (\ref{quarksleptons1}). The superpotential (\ref{spot3}) after the orbifold projection in the present case becomes 
\be 
W_{{\cal N}=1}^{(proj)}(h,f,f^c)=YTr(hf^cf),
\ee while the soft supersymmetry breaking terms are correspondingly read off from eq. (\ref{soft}). The SSB terms, discussed earlier in section 3.2, are still necessary in this context in order to generate a potential which leads to twisted fuzzy sphere vacua. 
However, 
in the present case the choice $h_{ijk}=\epsilon_{ijk}$ is not sufficient because the GUT and the EW breaking have to take place at different scales. We shall be more specific on this point when we study the most interesting case of $SU(3)^3$ in 5.3.

We would like to study the breaking of this model to the standard model, and
to see if this can be done using 
fuzzy orbifolds of the type (\ref{twisted-vacuum}). According to the rules of section 2, the surviving scalars and their 
superpartners live in
\be
\phi^i, \psi^i \in \left(\begin{array}{ccc}
0 & (4,2,1) & 0 \\
0 & 0 & (1,2,2) \\
(\bar 4,1,2) & 0 & 0\end{array}\right).
\ee

The first step is to break the $SU(4)\times SU(2)_R$ to $SU(3)_c$ 
(plus abelian factors),
which can be done with a $(\bar 4,1,2)$. 
Since we would like the hypercharge
\be
Y = B-L+2 T^3_R
\label{hypercharge-LR}
\ee
to survive \cite{Mohapatra:1986uf}, 
we need a vev which has zero hypercharge, as 
does $\nu^c_L$ in (\ref{quarksleptons1}). This is achieved by giving the following vev
\be
\langle\phi_{(\bar 4,1,2)}\rangle  =  \left(\begin{array}{cccc} 0 & 0 & 0 & 0
\\ 
0 & 0 & 0 & 1 \end{array}\right). 
\label{VEV-1}
\ee
This breaks the original nonabelian factors to 
$SU(3)_c \times SU(2)_L$, and three $U(1)$s 
survive. 
Let us note that $B-L$ is not preserved.

Next let us consider the electroweak (EW) breaking.
Since the electric charge is given by \cite{Mohapatra:1986uf}
\be
Q = T^3_L + \frac Y2 = T^3_L + T^3_R + \frac{B-L}2
\ee
this can be achieved by $\phi \in (1,2,2)$ provided
$[\sigma_3,\langle\phi\rangle] = 0$,
i.e. by a diagonal matrix
\be
\langle\phi_{(1,2,2)}\rangle =   \left(\begin{array}{cc} k & 0 
\\ 
0 & k' \end{array}\right). 
\ee
However, this procedure leads to extra $U(1)$ bosons having masses of the order of the electroweak scale instead of superheavy ones. 
Therefore this model does not seem to be phenomenologically viable. 

Nevertheless the following observation seems worthwhile. 
From the fuzzy orbifold point of view 
it is more natural to consider the breaking with the
following rank one fields
\bea
\langle\phi_{(1,2,2)}\rangle &=&   \left(\begin{array}{cc} 0 & 0  
\\ 
0 & k \end{array}\right) 
\qquad \mbox{and / or} \nn\\
\langle\phi_{(4,2,1)}\rangle &=& \left(\begin{array}{cc} 0 & 0 \\ 0 & 0 \\ 0 & 0 \\ 0 & k'
\end{array}\right), 
\label{EEW-Higgs-2}
\eea
both of which lead to the desired EW breaking. The point is that these
vevs satisfy the 
relations of a twisted fuzzy sphere.
Indeed, if we denote the vevs as
\be
H^1 = \langle \phi_{(4,2,1)}\rangle, \quad
H^2 = \langle \phi_{(1,2,2)}\rangle, 
\quad H^3 = \langle \phi_{(\bar 4,1,2)}\rangle,
\ee
(understood as sub-block matrices of the $8\times 8$ matrix
$\phi^i$),
we observe that 
\be
 H^1\, H^2 \sim
\obar{ H^3} \qquad \mbox{and cyclic permutations}
\ee 
This almost amounts to a fuzzy orbifold vacuum (\ref{twisted-vacuum}),
which is expected to arise in the presence
of the SSB potential (\ref{soft}).
Note that the different $H_i$ may well get different scales, 
which is necessary anyway as we noted above. In particular, if we denote the 
modulus of the vevs corresponding to the 
$|H_i|$ as $k_i$, then the commutation relations become 
\be
\,[H^i,H^j] = i h_{ijk} \obar H^k,
\ee
with 
\be h_{ijk}\equiv \frac{k_ik_j}{k_k}\e_{ijk}, \ee 
(no summation implied). 
This is a slight generalization of the twisted fuzzy sphere to include more than one scales, 
and it will be discussed further in section 5.3 in the most promising case of $SU(3)^3$.

\subsection{$SU(4)^3$ model}

This model can be obtained by starting initially with a $SU(12)$ gauge theory.
We perform an orbifold projection of the ${\cal N}=4$, $SU(12)$ SYM theory such that the $\g_3$ of eq. (\ref{g3}) takes the form
\be \g_3 = \left(\begin{array}{ccc} \one_{4} & 0 & 0 \\
        0 & \omega\one_{4} & 0 \\
        0 & 0 & \omega^2\one_{4} \end{array}\right). \ee
Then, according to the rules of section 2, the gauge group breaks down to $SU(4)^3$, with three chiral supermultiplets transforming as
\bea & SU(4)\times SU(4)\times SU(4) \nn\\
 & 3\cdot \biggl((4,\overline 4,1)+(\overline{4},1,4)+(1,4,\overline 4)\biggl). 
\eea
The quarks and leptons are accommodated as in eq. (\ref{quarksleptons2}). The superpotential (\ref{spot3}) after the orbifold projection becomes
\be 
W_{{\cal N}=1}^{(proj)}(h,f,f^c)=YTr(hf^cf),
\ee while the soft supersymmetry breaking terms are correspondingly read off from eq. (\ref{soft}).

The unification of quarks and leptons within $SU(4)_c$ leads to two possible formulae for the electric charge $Q$, merely due to the fact that the electric charges of the new heavy particles are not yet fixed. The two possibilities are \cite{Ma:2004mi} \bea Q_1&=&\frac 12 (B-L)+I_{3L}+I_{3R}, \nn\\ 
Q_2&=&\frac 12 (B-L)+I_{3L}+I_{3R}+I_{3L}'+I_{3R}'. \eea The quarks and leptons do not transform under $SU(2)_L'$ or $SU(2)_R'$ and therefore their electric charges are not affected.

The study of these two different charge assignments in \cite{Ma:2004mi} showed that there exist phenomenological obstacles in the viability of the model. In particular, when the electric charge is given by $Q_2$ the value of sin$^2\theta_W$ becomes 3/14, which is not realistic. In the $Q_1$ case sin$^2\theta_W$ becomes 3/8 as usual. However, the pathology arises at the level of the symmetry breaking down to the MSSM. This is achieved when the neutral scalar components of $f,f^c$ and $h$ acquire vevs. The study of this symmetry breaking reveals that there is an extra unwanted $U(1)$ surviving. Since this unbroken $U(1)$ couples to all particles, including the known quarks and leptons, this model cannot be viable phenomenologically.

\subsection{$SU(3)^3$ model}

This model is obtained from an $SU(9)$ gauge theory as follows.
We perform an orbifold projection of the ${\cal N}=4$, $SU(9)$ SYM theory such that the $\g_3$ of eq. (\ref{g3}) takes the form
\be \g_3 = \left(\begin{array}{ccc} \one_{3} & 0 & 0 \\
        0 & \omega\one_{3} & 0 \\
        0 & 0 & \omega^2\one_{3} \end{array}\right). \ee
Then, according to the rules of section 2, the gauge group breaks down to $SU(3)^3$, with three chiral supermultiplets transforming as
\bea & SU(3)\times SU(3)\times SU(3) \nn\\
 & 3\cdot \biggl((3,\overline 3,1)+(\overline{3},1,3)+(1,3,\overline 3)\biggl). 
\eea

First of all, the quarks of the first family transform under the gauge group as in eq. (\ref{quarks3}) and the leptons transform as in eq. (\ref{leptons3}). The superpotential (\ref{spot3}) after the orbifold projection in this case becomes \cite{Ma:2004mi}
\be 
W_{{\cal N}=1}^{(proj)}(\l,q,q^c)=YTr(\l q^cq)
+Y'\e_{ijk}\e_{abc}(\l_{ia}\l_{jb}\l_{kc}+q^c_{ia}q^c_{jb}q^c_{kc}
+q_{ia}q_{jb}q_{kc}),
\ee 
where the family superscripts are supressed.
The last terms are special in the $SU(3)^3$ case, and may involve different
families.
The soft supersymmetry breaking terms, which are necessary in order to obtain vacua in the form of twisted fuzzy spheres, are correspondingly read off from eq. (\ref{soft}) with the appropriate $h_{ijk}$ in order to incorporate different scales for the GUT and the EW symmetry breaking{\footnote{Of course the EW symmetry breaking of the MSSM requires the introduction of extra soft supersymmetry breaking terms (see e.g. \cite{Djouadi:2005gj}).}}. 

The spontaneous breaking of this unification model down to the MSSM has been studied in several publications \cite{Babu:1985gi,Ma:2004mi,Lazarides:1993uw} and it can be achieved in different ways. Here we would like to mention that in all the known symmetry breaking patterns either additional superfields have to be introduced in the theory \cite{Lazarides:1993uw} or the breaking has to happen in more than one steps, e.g. through the left-right symmetric model $SU(3)\times SU(2)_L\times SU(2)_R\times U(1)_{L+R}$ \cite{Ma:2004mi}.

Here we would like to present a different symmetry breaking pattern, where the initial $SU(3)^3$ gauge symmetry is spontaneously broken due to the existing scalar sector of the model, i.e. without the need of any additional superfields, and moreover the breaking happens in one step. In order to achieve this we shall utilize the fuzzy orbifold techniques which were presented previously.

Let us recall that the fields of one family can be represented by the following matrix,
\be\label{ql3} 
\left(\begin{array}{ccc}
0_3 & q & 0_{3}  \\
0_{3} & 0_3 & \l \\
q^c & 0_{3} & 0_3
\end{array}\right),
\ee where $0_3$ is the $3\times 3$ matrix with all the entries zero, or more explicitly as
\be 
\left(\begin{array}{ccccccccc}
0 & 0 & 0 & d&u&h&0&0&0 \\
0 & 0 & 0 & d&u&h&0&0&0\\
0 & 0 & 0 & d&u&h&0&0&0 \\
0 & 0 & 0 & 0&0&0 &N&E^c&\nu\\
0 &0&0 & 0 & 0 & 0 &E&N^c&e\\
0&0&0&0&0&0&\nu^c&e^c&S \\
d^c&d^c&d^c&0&0&0&0&0&0 \\
u^c&u^c&u^c&0&0&0&0&0&0 \\
h^c&h^c&h^c&0&0&0&0&0&0
\end{array}\right).
\ee 
Obviously the quark blocks cannot acquire a vev, since this would
break the colour $SU(3)$ gauge group factor. 
Therefore the term $Tr (\lambda q q^c)$ in the superpotential 
cannot play any role here.
The block which corresponds to the lepton supermultiplet may acquire
vevs only in the directions which have zero hypercharge. This means
that out of the nine components of this block only five may acquire a
vev, namely $S,\nu,\nu^c,N$ and $N^c$. The first three are responsible
for the breaking down to the MSSM, while the last two take care of the
EW breaking. Such a vacuum may indeed arise here
due to the presence of the 
$\e_{ijk}\e_{abc} \l_{ia}\l_{jb}\l_{kc}$ term in the
superpotential, and moreover we can
interpret it again in terms of a twisted fuzzy sphere.

To see the relation with a twisted fuzzy sphere
(\ref{twisted-vacuum}), we transform the 
lepton matrices as $\l'^i = \Omega_3 \lambda^i$ where 
$\Omega_3=
\left(\begin{array}{ccc}
0 & 1 & 0  \\
0 & 0 &1 \\
1 & 0 & 0 
\end{array}\right)$, noting that the relevant term 
$\e_{ijk}\e_{abc} \l_{ia}\l_{jb}\l_{kc}$ is invariant 
(up to sign) under such a
transformation. Then $\l$ is transformed to
\be \l'=
\left(\begin{array}{ccc}
E & N^c & e  \\
\nu^c & e^c & S \\
N & E^c & \nu 
\end{array}\right).\ee
Now consider a vacuum solution of the form (superscripts here denote families):
\bea \label{l1}
\l'^1&=&\left(\begin{array}{ccc}
0&k_1 & 0 \\ 0&0&0 \\0&0&0
\end{array}\right), \\ \label{l2}
\l'^2&=&\left(\begin{array}{ccc}
0 & 0&0\\ 0&0&k_2\\ 0&0&0
\end{array}\right),\\ \label{l3}
\l'^3&=&\left(\begin{array}{ccc}
0 & 0&0\\ 0&0&0\\ k_3&0&0
\end{array}\right),
\eea while everything else acquires a vanishing vev. These vevs
correspond to the directions of $N,N^c$ and $S$. The above matrices
satisfy 
\be [\l'^i,\l'^j]=ih_{ijk}(\l'^k)^{\dagger}, \ee where we have defined again
\be h_{ijk}\equiv \frac{k_ik_j}{k_k}\e_{ijk}. \ee This is a
generalization of the twisted fuzzy sphere vacuum where more than one
scales may be included. In the present model this is desirable, since
at least two scales have to be introduced, corresponding to the GUT
and EW breaking respectively. 
Moreover, since the model enjoys ${\cal N}=1$ supersymmetry these scales may in principle remain separate.

On the other hand, if we 
transform the 
lepton matrices as $\l''^i = \Omega'_3 \lambda^i$ where 
$\Omega'_3=
 \left(\begin{array}{ccc}
0 & 1 & 0  \\
 1 & 0 &0 \\
 0 & 0 & 1 
 \end{array}\right)$, then
\be \l''=
\left(\begin{array}{ccc}
E & N^c & e  \\
N & E^c & \nu \\
\nu^c & e^c & S 
\end{array}\right).
\ee 
The same twisted fuzzy sphere vacuum as before, namely the matrices (\ref{l1})-(\ref{l3}), corresponds now to the directions $\nu,\nu^c$ and $N^c$. Therefore, with the above procedure all the neutral directions acquire a vev and the original $SU(3)^3$ model is spontaneously broken down to $SU(3)_c\times U(1)_{em}$. In particular, at the scale where the directions $\nu,\nu^c$ and $S$ acquire vevs, $SU(3)^3$ is spontaneously broken down to the MSSM. Subsequently, at the scale where the $N$ and $N^c$ directions acquire vevs the breaking down to $SU(3)_c\times U(1)_{em}$ takes place. As we have already mentioned these scales are hopefully kept separate by supersymmetry. In other words the hierarchy problem is the same as in any supersymmetric particle physics model. 

The remarkable new result of the above procedure is that the spontaneous breaking of the $SU(3)^3$ model acquires an interesting geometrical explanation. It takes place solely due to the Higgsing of the twisted fuzzy spheres in the extra dimensions, without the need of any additional superfields and without the need of any intermediate breaking.

\section{Conclusions}

The proposal that more than four dimensions may exist in nature is a fruitful arena in modern theoretical physics in order to study the possibility to achieve unification of all the fundamental interactions. In an attempt to determine vacua of higher-dimensional unified theories which could lead to phenomenologically acceptable low-energy models several compactification schemes have been developed and a multitude of different manifolds describing the extra dimensions have been used.

In \cite{Aschieri:2005wm} it was proposed that the extra dimensions
can be described by matrix approximations of smooth manifolds,
i.e. fuzzy spaces. 
In spite of several remarkable features of this
approach, chiral fermions could not be accommodated so far 
in this framework. 
In an attempt to overcome this limitation, in \cite{Aschieri:2006uw} a
reverse approach motivated by the idea of deconstruction of dimensions
\cite{Arkani-Hamed:2001ca} was studied, where the fuzzy extra
dimensions are dynamically generated within a four-dimensional
renormalizable field theory. Nevertheless, only 
 mirror models could be obtained so far.

In the present paper, motivated by \cite{Kachru:1998ys}, we introduced an additional structure in the above context, based on orbifolds, in order to obtain chiral low-energy models. In particular we performed a $\Z_3$ orbifold projection of a ${\cal N}=4$ $SU(3N)$ SYM theory, which leads to a ${\cal N}=1$ supersymmetric theory with gauge group $SU(N)^3$. Adding a suitable set of soft supersymmetry breaking terms in the ${\cal N}=1$ theory, certain vacua of the theory were revealed, where twisted fuzzy spheres are dynamically generated. It is well known that the introduction of a soft supersymmetry breaking sector is not only natural but also necessary in the constructions of phenomenologically viable supersymmetric theories, with prime example the case of the MSSM \cite{Djouadi:2005gj}. Such vacua correspond to models which behave at intermediate energy scales as higher-dimensional theories with a finite Kaluza-Klein tower of massive modes and a chiral low-energy spectrum. The most interesting chiral models for low-energy phenomenology which can be constructed in this context turn out to be $SU(4)\times SU(2)\times SU(2), SU(4)^3$ and $SU(3)^3$.

Subsequently, the possibility to achieve further breaking of the above
models down to the MSSM and $SU(3)_c\times U(1)_{em}$ using twisted fuzzy
spheres was studied and it was shown that this is indeed
possible. Thus the spontaneous symmetry breaking of these unification
groups acquires an interesting geometrical explanation in terms of
twisted fuzzy spheres. 
The most interesting case is the trinification group $SU(3)^3$, 
which can be promoted even to an all-loop finite theory 
(for a review see \cite{Heinemeyer:2010xt}) and therefore 
it is suitable to make predictions \cite{Ma:2004mi,Heinemeyer:2009zs}. 

We have thus shown that fuzzy extra dimensions can arise in simple
field-theoretical models which are chiral, renormalizable, 
and may be phenomenologically viable. Moreover,
since some of these models can be 
finite with fermions in the adjoint of an underlying
$SU(3N)$ gauge group, these models can be  
generalized into the framework of Yang-Mills matrix model
such as \cite{Ishibashi:1996xs,Aoki:2002jt}.

\paragraph{Acknowledgements:} This work is supported by the NTUA programmes for basic research PEVE 2008 and 2009, the European Union's RTN programme under contract MRTN-CT-2006-035505 and the European Union's ITN programme "UNILHC" PITN-GA-2009-237920. The work of H.S. is supported in part by the FWF project P20017 and in part by the FWF project P21610.


\begin{thebibliography}{99}


\bibitem{Georgi:1974sy}
  H.~Georgi and S.~L.~Glashow,
  Phys.\ Rev.\ Lett.\  {\bf 32}, 438 (1974).

\bibitem{Pati:1973rp}
  J.~C.~Pati and A.~Salam,
  Phys.\ Rev.\ Lett.\  {\bf 31}, 661 (1973).

\bibitem{Georgi:1974yf}
  H.~Georgi, H.~R.~Quinn and S.~Weinberg,
  Phys.\ Rev.\ Lett.\  {\bf 33}, 451 (1974).

\bibitem{Fritzsch:1974nn}
  H.~Fritzsch and P.~Minkowski,
  Annals Phys.\  {\bf 93}, 193 (1975).

\bibitem{Amaldi:1991cn}
  U.~Amaldi, W.~de Boer and H.~Furstenau,
  Phys.\ Lett.\  B {\bf 260}, 447 (1991).

\bibitem{Dimopoulos:1981zb}
  S.~Dimopoulos and H.~Georgi,
  Nucl.\ Phys.\  B {\bf 193} (1981) 150.

\bibitem{Sakai:1981gr}
  N.~Sakai,
  Z.\ Phys.\  C {\bf 11}, 153 (1981).

\bibitem{Forgacs:1979zs}
  P.~Forgacs and N.~S.~Manton,
  Commun.\ Math.\ Phys.\  {\bf 72}, 15 (1980).

\bibitem{Kapetanakis:1992hf}
  D.~Kapetanakis and G.~Zoupanos,
  Phys.\ Rept.\  {\bf 219}, 1 (1992).

\bibitem{Kubyshin:1989vd}
  Yu.~A.~Kubyshin, I.~P.~Volobuev, J.~M.~Mourao and G.~Rudolph,
 Leipzig Univ. - KMU-NTZ-89-07 (89,REC.SEP.) 80p

\bibitem{Scherk:1979zr}
  J.~Scherk and J.~H.~Schwarz,
  Nucl.\ Phys.\  B {\bf 153}, 61 (1979).

\bibitem{Manton:1981es}
  N.~S.~Manton,
  Nucl.\ Phys.\  B {\bf 193} (1981) 502.
  G.~Chapline and R.~Slansky,
  Nucl.\ Phys.\  B {\bf 209}, 461 (1982).

\bibitem{Gross:1985fr}
  D.~J.~Gross, J.~A.~Harvey, E.~J.~Martinec and R.~Rohm,
  Nucl.\ Phys.\  B {\bf 256} (1985) 253.



\bibitem{Polchinski:1998rr}
  J.~Polchinski,
  ``String theory. Vol. 2: Superstring theory and beyond,''
{\it  Cambridge, UK: Univ. Pr. (1998) 531 p}


\bibitem{Aldazabal:2000sa}
  G.~Aldazabal, L.~E.~Ibanez, F.~Quevedo and A.~M.~Uranga,
  JHEP {\bf 0008} (2000) 002
  [arXiv:hep-th/0005067].

\bibitem{Candelas:1985en}
  P.~Candelas, G.~T.~Horowitz, A.~Strominger and E.~Witten,
  Nucl.\ Phys.\  B {\bf 258}, 46 (1985).


\bibitem{LopesCardoso:2002hd}
  G.~Lopes Cardoso, G.~Curio, G.~Dall'Agata, D.~Lust, P.~Manousselis and G.~Zoupanos,
  Nucl.\ Phys.\  B {\bf 652} (2003) 5
  [arXiv:hep-th/0211118].

\bibitem{Gauntlett:2003cy}
  J.~P.~Gauntlett, D.~Martelli and D.~Waldram,
  Phys.\ Rev.\  D {\bf 69}, 086002 (2004)
  [arXiv:hep-th/0302158].


\bibitem{Manousselis:2000aj}
  P.~Manousselis and G.~Zoupanos,
  Phys.\ Lett.\  B {\bf 504} (2001) 122
  [arXiv:hep-ph/0010141],
  Phys.\ Lett.\  B {\bf 518} (2001) 171
  [arXiv:hep-ph/0106033],
  JHEP {\bf 0203} (2002) 002
  [arXiv:hep-ph/0111125],
  JHEP {\bf 0411} (2004) 025
  [arXiv:hep-ph/0406207].


\bibitem{Maldacena:1997re}
  J.~M.~Maldacena,
  Adv.\ Theor.\ Math.\ Phys.\  {\bf 2} (1998) 231
  [Int.\ J.\ Theor.\ Phys.\  {\bf 38} (1999) 1113]
  [arXiv:hep-th/9711200].

\bibitem{Kachru:1998ys}
  S.~Kachru and E.~Silverstein,
  Phys.\ Rev.\ Lett.\  {\bf 80} (1998) 4855
  [arXiv:hep-th/9802183].

\bibitem{Douglas:1996sw}
  M.~R.~Douglas and G.~W.~Moore,
  ``D-branes, Quivers, and ALE Instantons,''
  arXiv:hep-th/9603167.

\bibitem{Douglas:1997de}
  M.~R.~Douglas, B.~R.~Greene and D.~R.~Morrison,
  Nucl.\ Phys.\  B {\bf 506} (1997) 84
  [arXiv:hep-th/9704151].


\bibitem{Aschieri:2005wm}
  P.~Aschieri, J.~Madore, P.~Manousselis and G.~Zoupanos,
  ``Renormalizable theories from fuzzy higher dimensions,''
  arXiv:hep-th/0503039;
  Fortsch.\ Phys.\  {\bf 52}, 718 (2004)
  [arXiv:hep-th/0401200];
  JHEP {\bf 0404}, 034 (2004)
  [arXiv:hep-th/0310072].

\bibitem{Arkani-Hamed:2001ca}
  N.~Arkani-Hamed, A.~G.~Cohen and H.~Georgi,
  Phys.\ Rev.\ Lett.\  {\bf 86} (2001) 4757
  [hep-th/0104005];
  Phys.\ Lett.\  B {\bf 513} (2001) 232
  [arXiv:hep-ph/0105239].

\bibitem{Aschieri:2006uw}
  P.~Aschieri, T.~Grammatikopoulos, H.~Steinacker and G.~Zoupanos,
  JHEP {\bf 0609} (2006) 026
  [arXiv:hep-th/0606021].
  P.~Aschieri, H.~Steinacker, J.~Madore, P.~Manousselis and G.~Zoupanos,
  ``Fuzzy Extra Dimensions: Dimensional Reduction, Dynamical Generation and
  Renormalizability,''
  arXiv:0704.2880 [hep-th];

\bibitem{Steinacker:2003sd}
  H.~Steinacker,
  Nucl.\ Phys.\  B {\bf 679} (2004) 66
  [arXiv:hep-th/0307075].

\bibitem{Steinacker:2007ay}
  H.~Steinacker and G.~Zoupanos,
  JHEP {\bf 0709} (2007) 017
  [arXiv:0706.0398 [hep-th]].

\bibitem{Chatzistavrakidis:2009ix}
  A.~Chatzistavrakidis, H.~Steinacker and G.~Zoupanos,
  arXiv:0909.5559 [hep-th].

\bibitem{Maalampi:1988va}
  J.~Maalampi and M.~Roos,
  Phys.\ Rept.\  {\bf 186} (1990) 53.

\bibitem{Grosse:2010zq}
  H.~Grosse, F.~Lizzi and H.~Steinacker,
  arXiv:1001.2703 [hep-th];
  H.~Steinacker,
  Nucl.\ Phys.\  B {\bf 810} (2009) 1
  [arXiv:0806.2032 [hep-th]].


\bibitem{Iso:2001mg}
  S.~Iso, Y.~Kimura, K.~Tanaka and K.~Wakatsuki,
  Nucl.\ Phys.\  B {\bf 604} (2001) 121
  [arXiv:hep-th/0101102].

\bibitem{Azuma:2004zq}
  T.~Azuma, S.~Bal, K.~Nagao and J.~Nishimura,
  JHEP {\bf 0405} (2004) 005
  [arXiv:hep-th/0401038].

\bibitem{Aoki:2002jt}
  H.~Aoki, S.~Iso and T.~Suyama,
  Nucl.\ Phys.\  B {\bf 634} (2002) 71
  [arXiv:hep-th/0203277].

\bibitem{Brink:1976bc}
  L.~Brink, J.~H.~Schwarz and J.~Scherk,
  Nucl.\ Phys.\  B {\bf 121} (1977) 77.
  F.~Gliozzi, J.~Scherk and D.~I.~Olive,
  Nucl.\ Phys.\  B {\bf 122} (1977) 253.

\bibitem{Bailin:1999nk}
  D.~Bailin and A.~Love,
  Phys.\ Rept.\  {\bf 315} (1999) 285.



\bibitem{Lawrence:1998ja}
  A.~E.~Lawrence, N.~Nekrasov and C.~Vafa,
  Nucl.\ Phys.\  B {\bf 533}, 199 (1998)
  [arXiv:hep-th/9803015].



\bibitem{Kiritsis:2003mc}
  E.~Kiritsis,
  Fortsch.\ Phys.\  {\bf 52} (2004) 200
  [Phys.\ Rept.\  {\bf 421} (2005\ ERRAT,429,121-122.2006) 105]
  [arXiv:hep-th/0310001].

\bibitem{Madore:1991bw}
  J.~Madore,
  Class.\ Quant.\ Grav.\  {\bf 9} (1992) 69.

\bibitem{Djouadi:2005gj}
  A.~Djouadi,
  Phys.\ Rept.\  {\bf 459} (2008) 1
  [arXiv:hep-ph/0503173].


\bibitem{Madore:2000en}
  J.~Madore, S.~Schraml, P.~Schupp and J.~Wess,
  Eur.\ Phys.\ J.\  C {\bf 16} (2000) 161
  [arXiv:hep-th/0001203].

\bibitem{Pati:1974yy}
  J.~C.~Pati and A.~Salam,
  Phys.\ Rev.\  D {\bf 10} (1974) 275
  [Erratum-ibid.\  D {\bf 11} (1975) 703].



\bibitem{Ibanez:1998xn}
  L.~E.~Ibanez,
  JHEP {\bf 9807} (1998) 002
  [arXiv:hep-th/9802103].

\bibitem{Ma:2004mi}
  E.~Ma, M.~Mondragon and G.~Zoupanos,
  JHEP {\bf 0412}, 026 (2004)
  [arXiv:hep-ph/0407236];

\bibitem{Glashow:1984gc}
  S.~L.~Glashow,
  ``Trinification Of All Elementary Particle Forces,''
 PRINT-84-0577-BOSTON-.

\bibitem{Rizov:1981dp}
  V.~A.~Rizov,
  Bulg.\ J.\ Phys.\  {\bf 8}, 461 (1981).


\bibitem{Lazarides:1993uw}
  G.~Lazarides and C.~Panagiotakopoulos,
  Phys.\ Lett.\  B {\bf 336}, 190 (1994)
  [arXiv:hep-ph/9403317].


\bibitem{Heinemeyer:2009zs}
  S.~Heinemeyer, E.~Ma, M.~Mondragon and G.~Zoupanos,
  arXiv:0910.0501 [hep-ph].


\bibitem{Babu:1985gi}
  K.~S.~Babu, X.~G.~He and S.~Pakvasa,
  Phys.\ Rev.\  D {\bf 33} (1986) 763.

\bibitem{Leontaris:2005ax}
  G.~K.~Leontaris and J.~Rizos,
  Phys.\ Lett.\  B {\bf 632} (2006) 710
  [arXiv:hep-ph/0510230].

\bibitem{Dienes:1998vh}
  K.~R.~Dienes, E.~Dudas and T.~Gherghetta,
  Phys.\ Lett.\  B {\bf 436} (1998) 55
  [arXiv:hep-ph/9803466].

\bibitem{Ghilencea:1998st}
  D.~Ghilencea and G.~G.~Ross,
  Phys.\ Lett.\  B {\bf 442} (1998) 165
  [arXiv:hep-ph/9809217].

\bibitem{Kobayashi:1998ye}
  T.~Kobayashi, J.~Kubo, M.~Mondragon and G.~Zoupanos,
  Nucl.\ Phys.\  B {\bf 550} (1999) 99
  [arXiv:hep-ph/9812221].

\bibitem{Kubo:1999ua}
  J.~Kubo, H.~Terao and G.~Zoupanos,
  Nucl.\ Phys.\  B {\bf 574} (2000) 495
  [arXiv:hep-ph/9910277].



\bibitem{Mohapatra:1986uf}
  R.~N.~Mohapatra,
  Berlin, Germany: Springer (1986) 309 P. (Contemporary Physics)


\bibitem{Heinemeyer:2010xt}
  S.~Heinemeyer, M.~Mondragon and G.~Zoupanos,
  arXiv:1001.0428 [hep-ph].


\bibitem{Ishibashi:1996xs}
 N.~Ishibashi, H.~Kawai, Y.~Kitazawa and A.~Tsuchiya,
 Nucl.\ Phys.\  B {\bf 498} (1997) 467
 [arXiv:hep-th/9612115].









\end{thebibliography}
\end{document}